\documentclass[12pt]{emulateapj}

\newcommand{\omc}{\hbox{$\omega$ Cen~}}
\newcommand{\omcp}{\hbox{$\omega$ Cen}}

\newcommand{\tuc}{\hbox{47~Tuc~}}
\newcommand{\tucp}{\hbox{47~Tuc}}

\def\Min{${}^{\prime}$\llap{.}}

\shorttitle{Relative distances of $\omega$ Centauri and 47~Tucanae } 
\shortauthors{Bono et al.}

\begin{document}
\title{On the relative distances of $\omega$ Centauri and 47~Tucanae\altaffilmark{1}}

\author{
G. \ Bono\altaffilmark{2,3},
P.\ B.\  Stetson \altaffilmark{4},
N. \ Sanna\altaffilmark{2,3},
A. \ Piersimoni\altaffilmark{5}, 
L.\ M.\  Freyhammer\altaffilmark{6},
Y. \ Bouzid\altaffilmark{7},
R. \ Buonanno\altaffilmark{3},
A. \ Calamida\altaffilmark{2},
F. \ Caputo\altaffilmark{2},
C. \ E. Corsi\altaffilmark{2},
A. \ Di Cecco\altaffilmark{2,3},
M. \ Dall'Ora\altaffilmark{8},
I. \ Ferraro\altaffilmark{2},
G. \ Iannicola\altaffilmark{2},
M. \ Monelli\altaffilmark{9},
M. \ Nonino\altaffilmark{10},
L. \ Pulone\altaffilmark{2},
C. \ Sterken\altaffilmark{7},
J. \ Storm\altaffilmark{11},
T. \ Tuvikene\altaffilmark{7}, and
A.\ R. \ Walker\altaffilmark{12}
}

\altaffiltext{1}
{Based on observations made with ESO telescopes SOFI@NTT, La Silla; 
ISAAC@VLT Paranal, projects: 66.D-0557, 68.D-0545, 075.D-0824, 
79.D-0765 and on images obtained from the ESO/ST-ECF. Optical 
data were also collected with the 1.0m SAAO telescope, projects: 
2006.1.012, 2006.2.02.}
\altaffiltext{2}{INAF--OAR, via Frascati 33, Monte Porzio Catone, Rome, Italy; bono@mporzio.astro.it.  
}
 \altaffiltext{3}{Dipartimento di Fisica, UniToV, via della Ricerca Scientifica 1, 00133 Rome, Italy}
 \altaffiltext{4}{DAO--HIA, NRC, 5071 West Saanich Road, 
      Victoria, BC V9E 2E7, Canada}
 \altaffiltext{5}{INAF--OACTe, via M. Maggini, 64100 Teramo, Italy}
 \altaffiltext{6}{Centre for Astrophysics, University of Central Lancashire, Preston PR1 2HE}
 \altaffiltext{7}{VUB, OBSS/WE, Pleinlaan 2, 1050 Brussels, Belgium}
 \altaffiltext{8}{INAF--OACN, via Moiariello 16, 80131 Napoli} 
 \altaffiltext{9}{IAC, Calle Via Lactea, E38200 La Laguna, Tenerife, Spain}
 \altaffiltext{10}{INAF--OAT, via G.B. Tiepolo 11, 40131 Trieste, Italy}
 \altaffiltext{11}{AIP, Sternwarte 16, 14482 Potsdam, Germany} 
 \altaffiltext{12}{CTIAO--NOAO, Casilla 603, La Serena, Chile}

\date{\centering drafted \today\ / Received / Accepted }

\begin{abstract}
We present precise optical and near-infrared ground-based photometry of 
two Globular Clusters (GCs): $\omega$ Cen and 47~Tuc. These 
photometric catalogs are unbiased in the Red Giant Branch (RGB) 
region close to the tip. We provide new estimates of the RGB tip 
(TRGB) magnitudes---$m_I(TRGB)=9.84\pm0.05$, $\omega$ Cen; 
$m_I(TRGB)=9.46\pm0.06$, 47~Tuc---and use these to determine the 
relative distances of the two GCs. We find that distance ratios
based on different calibrations of the TRGB, the RR Lyrae stars and 
kinematic distances agree with each other within one sigma. Absolute 
TRGB and RR Lyrae distance moduli agree within 0.10--0.15 mag, while 
absolute kinematic distance moduli are 0.2--0.3~mag smaller. 
Absolute distances to \tuc based on the Zero-Age-Horizontal-Branch
and on the white dwarf fitting agree within 0.1 mag, but they are
0.1--0.3 mag smaller than TRGB and RR Lyrae distances. 
\end{abstract}

\keywords{globular clusters: $\omega$ Centauri, 47~Tucanae}


\newpage 
\section{Introduction}

The road to quantitative astrophysics is paved with improvements in the 
absolute cosmic distance scale.  Accurate primary distances based on 
trigonometric parallaxes are limited to measurements with the 
HIPPARCOS satellite (van Leeuwen et al.\ 2007), with the Fine Guidance 
Sensor on board the Hubble Space Telescope (Benedict et al.\ 2007; 
Evans et al.\ 2008) and with near-infrared interferometers on the 
ground (Kervella et al. 2006). Secondary distance estimates have relied
either on period-luminosity relations for classical Cepheids (Sandage \& 
Tammann 2006;  Fouqu\'e, et al.\ 2007; Caputo 2008), RR Lyrae stars 
(Szewczyk et al.\ 2008), type II Cepheids 
(Feast et al.\ 2008; Groenewegen et al.\ 2008); 
or on fits to the main sequence 
(Gratton et al.\ 2003), the white dwarf (WD) cooling sequence 
(Zoccali et al.\ 2001; Layden et al.\ 2005), the Zero Age Horizontal Branch 
(ZAHB, Caloi \& D'Antona 2005; Salaris et al.\ 2007), or the TRGB 
(Madore \& Freedman 1995; Salaris \& Cassisi 1998; Bellazzini 2008). 
Kinematic distances to GCs offer promise as a new and 
independent {\it primary\/} distance indicator. This method is 
based on the ratio between the dispersions in proper motion and radial 
velocity of cluster members.  The accuracy of this method 
is limited only by the precision of the measurements and the size of the
sample and, barring urecognized systematic errors,
absolute distances with an accuracy of 1\% might be provided 
(King \& Anderson 2002). 

All the above methods have practical limitations, and a definitive 
consensus on low- and intermediate-mass distance indicators has not 
yet been reached. The typical problems affecting current distance 
estimates might be split into two groups:  {\em i)}-- precision, i.e., 
the ongoing effort to reduce random observational errors; and {\em ii)}-- 
accuracy, i.e., the neverending struggle to identify and eliminate systematic 
errors.  Problems associated with the former group can be 
alleviated by increases in sample size and in signal-to-noise 
ratio.  Usually, problems connected with the latter group can be 
recognized and addressed only by comparing results obtained by completely
different methodologies.
The number of GCs for which we have distance estimates based 
on different methods is quite limited. However, a few GCs do have distances
based on multiple methods including the TRGB (Bellazzini et al.\
2004), RR Lyraes (Del Principe et al.\ 2006, DP06; 
Sollima et al.\ 2008), and kinematics (van de Ven et al.\ 2006; 
McLaughlin et al.\ 2006). Even with this extremely limited sample, we have 
evidence that the kinematic distances are significantly and systematically smaller 
than distances based on the other methods. The difference 
ranges from 5\% for M15 (McNamara et al.\ 2004), to 10--15\% for both 
\omc (van de Ven et al.\ 2006; DP06) and 47~Tuc (McLaughlin et al.\ 2006). 
On the other hand, distances to \tuc based on the ZAHB and on the WD
fitting are on average 10\% smaller than distances based on other methods
(Salaris et al.\ 2007). 

\section{Observations and data reduction}

The $B$, $V$, and $I$ data considered for this investigation come from the
database of original and archival observations which have been collected,
reduced, and calibrated by one of the authors in an ongoing effort to provide
homogeneous photometry on the Landolt (1992) photometric system for a
significant corpus of astronomically interesting targets---including, most
particularly, globular clusters (Stetson 2000).  For \omcp, we currently have a
catalog consisting of 518,905 stars with at least two measurements in each
of the three optical bands; these span an area with extreme dimensions on the
sky of 37\Min2 (E-W) by 40\Min6 (N-S).  These data were obtained in the course
of 14 observing runs with various telescopes and cameras.  The corresponding
numbers for 47~Tuc are 193,679 stars in an area of 35\Min3 by 41\Min3 from 16
observing runs.  As we have said above, systematic errors cannot be found by
consideration of one data set or one methodology alone. However, in the absence 
of truly independent confirmation we nevertheless feel confident, on the basis
of the observed consistency of the results within and between the different
observing runs available to us, that these optical catalogs of the two clusters
are on a common photometric system to well under 0.01$\,$mag.  
 
To provide photometric indices with greater temperature sensitivity we
matched the optical catalog for \omc\ with a multiband NIR catalog derived
from observations obtained with ISAAC@VLT (ESO, Paranal) and SOFI@NTT (ESO, La
Silla). A significant fraction of the SOFI data have already been presented by
DP06, while the ISAAC data will be presented by Calamida
et al.\ (2008, in preparation). These data cover an area of $12\times12$ arcmin
centered on the cluster; we have supplemented them with 2MASS 
data for the more external regions. We ended up with a final catalog including 
$\sim 200,000$ stars with at least one measurement in at least two 
NIR bands. 
We ended up with a catalog including $\sim 198,000$ stars with at least one 
measurement in both optical and NIR bands.   

\section{Results and discussion}

To identify of the TRGB, we need to distinguish the
Asymptotic Giant Branch (AGB) stars.  To do this, we adopted optical--NIR CMDs. 
Data for \omc (see top panel of Fig.~1) show that in the $J$, $B-K$ CMD 
the AGB stars separate well from RGB stars for $11 \le J \le 12$, while at the 
bright end they are, at fixed color, slightly brighter than RGB stars.  

\begin{figure}[!ht]
\begin{center}
\label{fig1}
\includegraphics[height=0.50\textheight,width=0.45\textwidth]{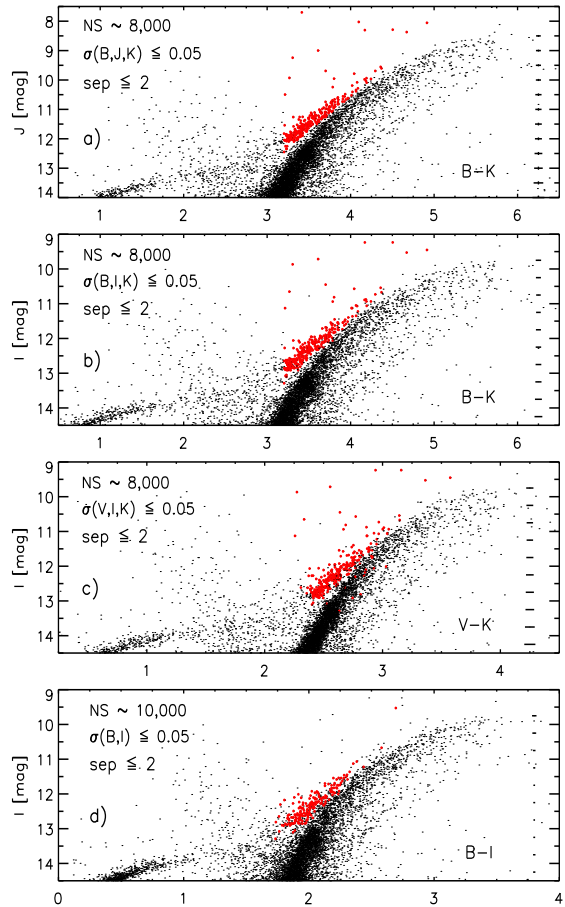}
\caption{{\em a)}-- Optical--NIR $J$, $B-K$ CMD of the bright region $J \le 14$ 
of \omcp. Red circles display candidate 
AGB stars. The number of stars selected and the adopted selection criteria 
(intrinsic photometric error, {\em separation}) are also labeled. The error
bars on the right display standard errors in magnitude and color.  
{\em b)}-- Same as the top, but for the $I$, $B-K$ CMD.
{\em c)}-- Same as the top, but for the $I$, $V-K$ CMD.
{\em d)}-- Same as the top, but for the $I$, $B-I$ CMD.
The red circles plotted in the last three panels are the candidate 
AGB stars selected from the $J$, $B-K$ CMD.
}
\end{center}
\end{figure}

The other panels show the distribution of bright stars in three different
optical/NIR CMDs for \omcp, where the red circles represent the candidate AGB stars
selected in the top panel. 
Overall, this figure suggests that the sample of bright RG stars is minimally 
contaminated by AGB stars.  Given that, we were able to estimate the RGB 
luminosity function. Note that the our sample includes $\sim$220 stars within one
$I$-band magnitude of the tip. This is roughly a factor of two larger than 
the number considered necessary for a robust TRGB detection (Madore \&
Freedman 1995) and $\approx$20\% larger than the sample adopted by Bellazzini
et al.\ (2001).

Data plotted in the top panel of Fig.~2 show a well defined jump in the star
counts for $m_I\sim 9.85$, which we take to  mark the position of the TRGB.  The
identification is supported by the smoothed luminosity function obtained using a
Gaussian kernel with standard deviation equal to the photometric error (Sakai et
al. 1996; middle panel). The bottom panel shows the response of the edge
detector, a four-point Sobel filter convolved with the smoothed luminosity
function; the dashed vertical lines mark the detection of the TRGB at
$m_I(TRGB)\sim 9.84\pm0.05$. It is worth noting that the current TRGB estimate
is in excellent agreement with the value provided by Bellazzini et al.\ (2001),
i.e., $m_I(TRGB)=9.84\pm0.05$.  To estimate the dereddened magnitude, 
we adopted a reddening of $E(B-V)=0.11\pm0.02$ (Kaluzny et al.\ 2002; 
Calamida et al.\ 2005).  This value, combined with the analytical 
relation for stellar extinction by Cardelli et al.\ (1989), gives for 
the Cousins $I$ passband: 
$A_I=0.59\times A_V$ = $0.59\times3.1\times0.11\pm0.02$=$0.20\pm0.06$ mag. 
Therefore, we end up with $m_{I,0}(TRGB)$=$9.64\pm0.08$. The error budget 
includes the uncertainties in the absolute zero-point calibration 
($< 0.01$), the cluster reddening, and the reddening law (Fitzpatrick 1999).

\begin{figure}[!ht]
\begin{center}
\label{fig2}
\includegraphics[height=0.30\textheight,width=0.45\textwidth]{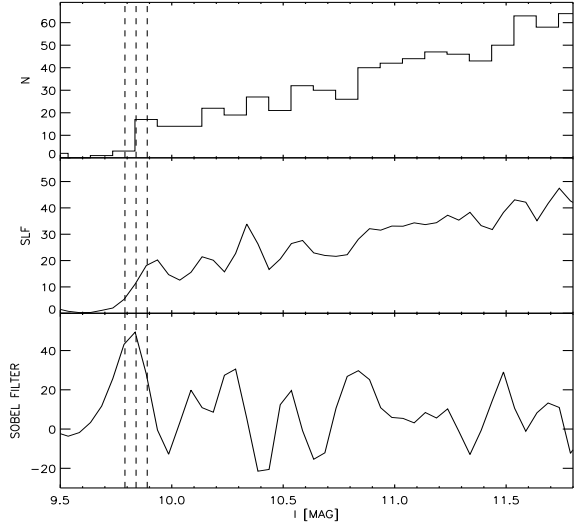}
\vspace*{-0.05truecm}
\caption{Top -- $I$-band luminosity function for the bright end of the RGB in
\omcp.  The vertical dashed lines indicate the position of the TRGB. 
Middle -- Smoothed luminosity function of the same RGB region obtained using a
Gaussian kernel with standard deviation equal to the photometric error (Sakai et
al.\ 1996). 
Bottom -- Response of the Sobel filter to the smoothed luminosity function.  
}
\end{center}
\end{figure}

We adopted the same approach to estimate the TRGB in 47~Tuc.  In the absence of
other NIR observations, we have cross-correlated our optical catalog with the
2MASS NIR catalog, resulting in a catalog including $\sim 15,000$ 
stars with measurements in optical and NIR bands. As illustrated by the 
error bars in Fig.~3, the photometric accuracy is very good 
from the TRGB down to the red HB ($H$ $\sim$ 12, $B-K$$\sim$3).  As was the case for
\omcp, we find that star counts close to the TRGB should be minimally contaminated 
by AGB stars.  
 
\begin{figure}[!ht]
\begin{center}
\label{fig3}
\includegraphics[height=0.50\textheight,width=0.45\textwidth]{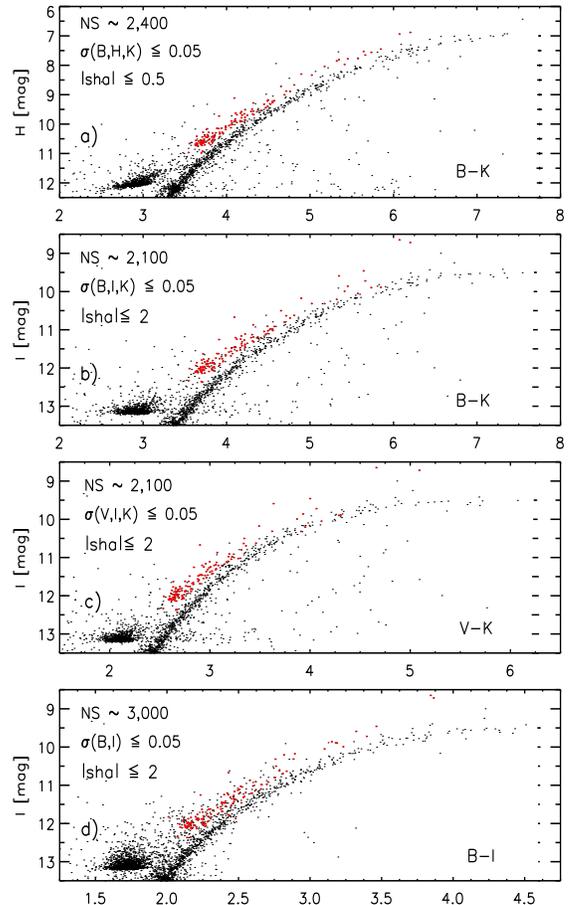}
\vspace*{-0.05truecm}
\caption{Same as Fig.~1, but the top panel shows the $H$, $B-K$ CMD of 
bright stars in 47~Tuc. Stars were selected according to intrinsic  
photometric error and {\em sharpness}.   
}
\end{center}
\end{figure}

We estimate that the TRGB in 47~Tuc is located at $m_I(TRGB)=9.46\pm0.06$
(see vertical lines in Fig.~4).  This estimate agrees quite well with the 
estimate of Bellazzini et al.\ (2004), i.e.,
$m_I(TRGB)=9.40\pm0.13$.  Note that the latter estimate includes a small
correction to allow for the limited number of RGs in the last magnitude
($N_{RGB}$=80).  Our sample includes 97 RGs in the same magnitude range---
an increase of $\sim$25\%---bringing us closer to the number 100 stars 
which was proposed as the reference density for the method.
We have therefore chosen not to apply this correction.  To estimate the
unreddened magnitude, we adopted $E(B-V)=0.04\pm0.02$ (Salaris et al.\ 2007) 
which, with the Cardelli et al. relation 
$A_I = 0.59\times A_V= 0.59\times3.1\times0.04\pm0.02=0.07\pm0.06$ mag, 
gives $m_{I,0}(TRGB)=9.39\pm0.08$. The error budget includes the same
sources of uncertainty as before.

\begin{figure}[!ht]
\begin{center}
\label{fig4}
\includegraphics[height=0.30\textheight,width=0.45\textwidth]{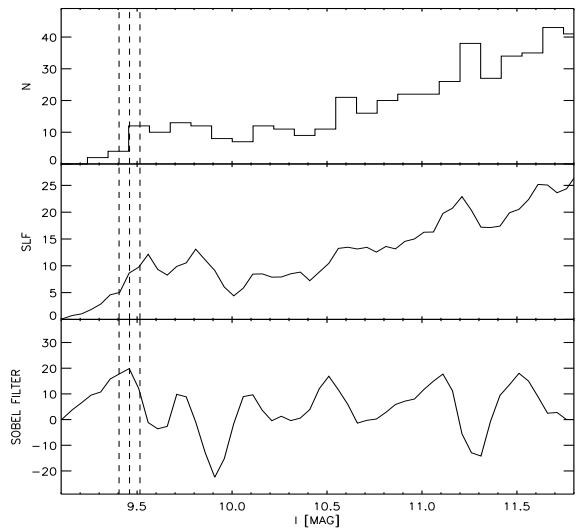}
\vspace*{-0.05truecm}
\caption{Same as Fig.~2, but for bright RG stars in 47~Tuc. 
}
\end{center}
\end{figure}

To use the TRGB method to estimate the relative distance independently, we must
correct for the difference in the clusters' metal content.  The TRGB calibration
provided by Salaris \& Cassisi (1998) relies on theoretical models, while that
provided by Bellazzini et al.\ (2004) relies on the absolute distance of \omc
based on the eclipsing binary ($13.65\pm0.11$, $E(B-V)=0.13\pm0.02$, Thompson et
al.\ 2001) and on an average distance for 47~Tuc ($13.31\pm0.11$, E(B-V)=0.04). 
For our purposes we require only the slope of the metallicity-TRGB relation, and
in this the two calibrations are in substantial agreement: using the quoted
relations and iron abundances of [Fe/H]=-1.7 (\omcp, metal-poor peak, 
Johnson et al.\ 2008) and [Fe/H]=-0.7 (47~Tuc), the differential corrections 
are +0.09 and +0.06 with an uncertainty of 0.03 if we assume a generous error 
allowance of 0.2 dex on the metallicity difference between the two clusters.
Therefore, splitting the difference between the two corrections, the relative
distance becomes $\delta \mu=9.64\pm0.08$(\omcp) -  $9.39\pm0.08$ (47~Tuc) +
$0.08 = 0.33\pm0.11$ mag.  For comparison, absolute kinematic distances to \omc 
and 47~Tuc have been recently provided by van den Ven et al.\ (2006) and 
by McLaughlin et al.\ (2006). The implied relative distance between the 
two clusters is 
$\delta \mu = 13.41\pm0.13$(\omcp) - $13.02\pm0.19$(47~Tuc) = $0.39\pm0.23$ mag. 

A similar value results if we use RR Lyrae stars to estimate the relative
distances: the variable V9 ($\log P=-0.1326$) is the only RR Lyrae currently
known in 47~Tuc and it is suspected to be slightly evolved (Storm et al.\ 1994). 
$K$-band magnitudes have been shown to be minimally affected by evolutionary
corrections (Bono et al.\ 2003; Catelan et al.\ 2004) and by assuming
$E(B-V)=0.04$ and the Cardelli law, we found $<k_0(V9)>=12.66\pm0.02$ mag. 
Fortunately, there are five RR Lyrae stars: V90, V127, V136, V87 (Kaluzny et al.\
2004) and V141 (Clement et al.\ 2001) in \omc with
periods ranging from $\log P= -0.155$ to $-0.119$ and with accurate $K$-band
mean magnitudes ($<K>=12.92\pm0.03$). In order to account for the difference in
metal abundances we adopted the metallicity estimates for individual RR Lyraes
provided by Rey et al.\ (2000) and found that the mean metallicity for these
five is $<[Fe/H]>=-1.53\pm0.17$.  The difference in mean magnitude as a function
of the iron content was estimated using the theoretical calibration 
of DP06. The independent theoretical calibration provided
by  Catelan et al.\ (2004) provides very similar distances (DP06). 
For a difference in metal abundance of 0.8 dex\footnote{Note that
the mean metallicity of the five RR Lyrae stars in \omc is not exactly the 
same as the RG metallicity that we employed for the TRGB.} 
the difference in the $K$ band is 0.07\footnote{Estimates based on the 
$K$-band P-L relations for Z=0.0001--0.001--0.004 listed in Table~1 of DP06} mag. 
Therefore, the relative distance based on RR Lyraes is:
$\delta \mu=  12.92\pm0.03$(\omcp) - $12.66\pm0.02$(47~Tuc) + 0.07 =
$0.33\pm0.07$~mag.  We thus find that {\it relative\/} distances based on three
completely independent methods agree within one sigma, or $\sim 3$\% in
distance, although admittedly the current analysis is based on only two GCs.   

We now want take a closer look at the consistency of our relative cluster
distances with inferences from various proposed absolute distance scales that
have been established on the basis of much larger samples of GCs.  The
zero-point of the TRGB calibration provided by Lee et al.\ (1993) relies on
cluster distances obtained with the $M_V$ vs [Fe/H] relation for RR Lyrae stars. 
On this distance scale \omc $\mu = (9.64\pm0.08) + (4.01\pm0.05)=13.65\pm0.09$, and
for 47~Tuc $\mu = (9.39\pm0.08) + (3.93\pm0.05)=13.32\pm0.09$ mag, for a modulus
difference of $\delta \mu = 0.33\pm0.13$ mag.  We did not consider the TRGB
calibration provided by Rizzi et al.\ (2007), since 47~Tuc is outside the
metallicity range covered by this calibration. The use of the empirical 
color-metallicity relation for TRGB stars by Bellazzini et al.\ (2001) 
and of the Rizzi's TRGB calibration would provide distances very similar 
to the above ones. The TRGB calibration provided by Salaris \& Cassisi (1998) 
is brighter in absolute terms, but new findings (Cassisi et al.\ 2007) 
indicate that the discrepancy is significantly reduced.  

From absolute distances based on the empirical $K$-band P-L relation for RR Lyrae
stars provided by Sollima et al.\ (2008) we found $\mu = 13.75\pm0.11$ for \omc
while for 47~Tuc we found, using the same unreddened magnitude as before for V9,
$\mu = 13.47\pm0.11$, and therefore $\delta \mu = 0.28\pm0.14$ mag.  
We note that the quoted absolute distances to \omc agree quite well with the
distances based on the eclipsing binary ($13.71\pm0.11$,
$13.75\pm0.04$, $E(B-V)=0.11\pm0.02$, Thompson et al.\ 2001; Kaluzny et al.\
2002). The distances to 47~Tuc agree quite well with the distance based on main
sequence fitting ($13.40\pm0.03$, E(B-V)=0.04, Gratton et al.\ 2003), but are
systematically larger than distances based on fitting the WD cooling
sequence ($13.15\pm0.14$, E(B-V)=0.04, Zoccali et al.\ 2001) and the ZAHB
($13.18\pm0.03$, Salaris et al.\ 2007). Unfortunately, robust distance estimates
to \omc based on these two methods are either not available or hampered by its
spread in metal content.  However, the use of these moduli and of the average
TRGB and RR Lyrae relative distances would provide distances to \omc
similar to the kinematic distances, but systematically smaller than the 
other standard candles.   

Thus, we are left with the following evidence: {\em i)} differential
distance moduli or---equivalently---distance ratios based upon (a)~the
absolute $I$-band relation for the TRGB, (b)~the $K$-band P-L relations for RR Lyrae stars,
and (c)~the kinematic method agree with each other to within the precision of each
method, or roughly 0.06~mag (see Table~1).  This finding suggests
that the estimated metallicity dependences of the different methods are
consistent among themselves, since these two clusters differ in metallicity
by $\sim 1$~dex.  However, we obviously cannot completely exclude the possibility that
the agreement is either fortuitous or a result of error cancellation. 
{\em ii)} Absolute distances based on current calibrations of
the TRGB and RR Lyrae methods agree with each other to 0.10--0.15~mag, 
but the kinematic distance scale is 0.2--0.3~mag shorter (see Table~1). 
The same outcome applies to \tuc distances based on ZAHB and WD fitting,
they agree with each other within 0.10 mag, but they are 0.1--0.3 mag
smaller than TRGB and RR Lyrae distances (Salaris et al.\ 2007).
Taken together, points {\em i)} and
{\em ii)} suggest that our Population~II distance scale is {\it not\/}
at present limited by random errors due to data precision or sample size, but
rather by systematics:  either a systematic error shared by the TRGB method and
the RR Lyrae method, due perhaps to a common rung in the calibration ladder that
leads to their respective zero points, or a systematic bias inherent to the
kinematic method that is common to the independent studies of \omc\ and 47~Tuc.  

While we wait for accurate trigonometric parallaxes for GCs from SIM and GAIA,
similar analyses of new GCs should help to clarify the relative impact of 
random and systematic errors among the various distance indicators.




\tablewidth{0pt}                       
\begin{deluxetable}{cccccc}
\scriptsize
\tablecaption{Relative and absolute distance moduli (mag) to \omc and \tucp.}
\tablehead{
\colhead{}&
\colhead{TRGB\tablenotemark{a}}&
\colhead{RR Lyrae\tablenotemark{b}}&
\colhead{Cl. Kin.\tablenotemark{c}}&
\colhead{TRGB\tablenotemark{d}}&
\colhead{RR Lyrae\tablenotemark{e}}
}
\startdata
\omcp         & $9.64\pm0.08$ & $12.92\pm0.03$ & $13.41\pm0.13$ & $13.65\pm0.09$ & $13.75\pm0.11$\\
\tucp         & $9.39\pm0.08$ & $12.66\pm0.02$ & $13.02\pm0.19$ & $13.32\pm0.09$ & $13.47\pm0.11$\\
$\delta \mu$  & $0.33\pm0.11$ & $0.33\pm0.07$  & $0.39\pm0.23$  & $0.33\pm 0.13$ & $ 0.28\pm0.14$\\
\enddata
\tablenotetext{a}{Distances based on $m_{I,0}(TRGB)$.}
\tablenotetext{b}{Distances based on $m_{K,0}(RR Lyrae)$.}  
\tablenotetext{c}{Distances based on cluster kinematics.}  
\tablenotetext{d}{Distances based on the $M_{I}(TRGB)$ calibration by Lee et al.\ (1993).}
\tablenotetext{e}{Distances based on the $M_{K}(RR Lyrae)$ calibration by Sollima et al.\ (2008).}
\end{deluxetable}

\end{document}